# Emerging Oscillating Reactions at the Insulator/Semiconductor Solid/Solid Interface via Proton Implantation

Dechao Meng[1,2*], Guanghui Zhang[1,2], Ming Li[2], Zeng-hui Yang[1,2*], Hang Zhou[1,2], Mu Lan[1,2], Yang Liu[1,2], Shouliang Hu[2], Yu Song[3], Chunsheng Jiang[1,2], Lei Chen[2], Hengli Duan[4], Wensheng Yan[4], Jianming Xue[5], Xu Zuo[6], Yijia Du[1,2], Gang Dai[1,2*], Su-Huai Wei[7*]

[1]Microsystem and Terahertz Research Center, China Academy of Engineering Physics, Chengdu 610200, P. R. China
[2]Institute of Electronic Engineering, China Academy of Engineering Physics, Mianyang 621999, P. R. China
[3]College of Physics and Electronic Information Engineering, Neijiang Normal University, Neijiang 641112, China
[4]National Synchrotron Radiation Laboratory, University of Science and Technology of China, Hefei, 230026, P. R. China
[5]State Key Laboratory of Nuclear Physics and Technology, Peking University, Beijing 100193, P. R. China.
[6]College of Electronic Information and Optical Engineering, Nankai University, Tianjin 300071, China.
[7]Beijing Computational Science Research Center, Beijing 100193, P. R. China.

*Corresponding author:

mengdechao_mtrc@caep.cn, yangzenghui_mtrc@caep.cn, daigang_mtrc@caep.cn, suhuaiwei@csrc.ac.cn

# Abstract

Most oscillating reactions (ORs) happen in solutions. Few existing solid-based ORs either happen on solid/gas (e.g., oxidation or corrosion) or solid/liquid interfaces, or at the all-solid interfaces neighboring to metals or ionic conductors (e.g., electrolysis or electroplate). We report in this paper a new type of all-solid based OR that happens at the insulator (amorphous $SiO_2$)/semiconductor (Si) interface with the interfacial point defects as the oscillating species. This OR is the first example of the point-defect coupled ORs (PDC-ORs) proposed by H. Schmalzried et al. [1] and J. Janek et al. [2] decades ago. We use proton implantation as the driving force of the oscillation, and employ techniques common in semiconductor device characterization to monitor the oscillation *in situ*. This approach not only overcomes the difficulties associated with detecting reactions in solids, but also accurately measure the oscillating ultra-low concentration ($10^{10}$~$10^{11}$ cm$^{-2}$) of the interfacial charged point-defects. We propose a mechanism for the reported PDC-OR based on the Brusselator model by identifying the interfacial reactions.

# I Introduction

Oscillating reactions (ORs) [1-16] is a special and fascinating type of reactions: the concentrations of species in regular chemical reactions change monotonically in time, but in ORs the concentrations of intermediates would oscillate in time, and in some cases the oscillation happens in space as well [4, 6, 8, 13]. It is a manifestation of the Nobel-prize-winning study of the symmetry-breaking instabilities by Prigogine et al. [17,18] Such behaviors may occur only when the system is far from equilibrium, and the precise conditions can be derived in principle [17-19].

ORs in solutions usually have complicated mechanisms. For example, the Field-Koros-Noyes (FKN) mechanism [20] of the Belousov-Zhabotinsky (BZ) reaction [21, 22] contains 21 intermediate species and 18 elementary reactions. Prigogine et al. proposed commonly acknowledged simplified models of ORs in solutions ( Brusselators) [17-19], and the simplest model is

$$A \rightarrow X, \qquad (B1)$$

$$2X+Y \rightarrow 3X, \qquad (B2)$$

$$B+X \rightarrow Y+D, \qquad (B3)$$

$$X \rightarrow E, \qquad (B4)$$

where A, B are reactants, D, E are products, and X, Y are the intermediates. The rate equations of X and Y are

$$\frac{d}{dt}[X] = [A] + [X]^2[Y] - [B][X] - [X], \qquad (1)$$

$$\frac{d}{dt}[Y] = -[X]^2[Y] + [B][X], \qquad (2)$$

where the rate constants are set to 1 for simplicity, and the square brackets denote concentrations. The auto-catalytic nature of the reactions leads to both positive and negative terms (feedbacks) [2, 6, 19] containing the concentrations of the intermediates in the rate equations of these intermediates, which is essential for non-linear temporal or spatial behaviors of concentrations.

While ORs typically happen in liquid (aqueous solutions) [21-27], a few chemical oscillating phenomena (refer to Fig. 1) have been reported in homogenous soft matter [13] and more at the inhomogeneous interfaces (soft matter/liquid interface [28-31], solid-gas [14, 32-33], solid-solid [1-2, 34-36] and liquid/solid coupled with liquid/liquid nicknamed as "Beating Heart" [16, 37-38]) They may happen either due to the higher mobility of reactants in liquid phases [21-31] or due to interfacial matter transport facilitated by external driving forces in solid [2, 14, 32-36]. The known all-solid-based ORs have external stimuli (temperature [14, 32-36], electrical field [2]) acting as driving forces, with thermal [32] and mechanical [33] factors as negative feedbacks. The all-solid-based ORs involves fewer reactions than those in solutions and the neighboring solids are metal or solid ionic conductor. None of the known solid-solid interfacial ORs involve insulators or semiconductors.

Schmalzried et al. [1] proposed in 1995 the possibility of a point-defect coupled mechanism for oscillations at the solid-solid interface, where the cross-interface transport of point defects and coupled interfacial relaxation upset the local equilibrium and induce oscillations. Janek et al. [2] further expanded this proposal by pointing out that such point-defect coupled ORs (PDC-ORs) may be achieved following the creation of instabilities (such as vacancies) due to atoms/ions transfer across the interface. Nevertheless, none PDC-ORs have been found yet as far as we know. Lacking suitable *in situ* experimental methods for determining the properties of point defects and the interface makes it even harder to find PDC-ORs.

In this paper, we report a new OR at the interface between amorphous $SiO_2$ and Si (a-$SiO_2$/Si, 'a' stands for amorphous) that (1) is the first direct evidence of PDC-OR and (2) is the first OR that happens at the interface between an insulator and a semiconductor. (3) The oscillation is made possible by proton implantation, which has never been employed before as the external driving force in ORs in solids. (4) We circumvent the difficulty of *in situ* monitoring the point defects inside a solid with techniques common in the electrical characterization of semiconductor devices. (5) We find that certain characteristics of the new OR in solid resembles that of ORs in solutions, and we propose a Brusellator-like model as the first attempt at explaining the mechanism. Fig. 1 reviews the known ORs and puts the reaction reported in this work into perspective, and more details are available in the Supplemental Information (SI).

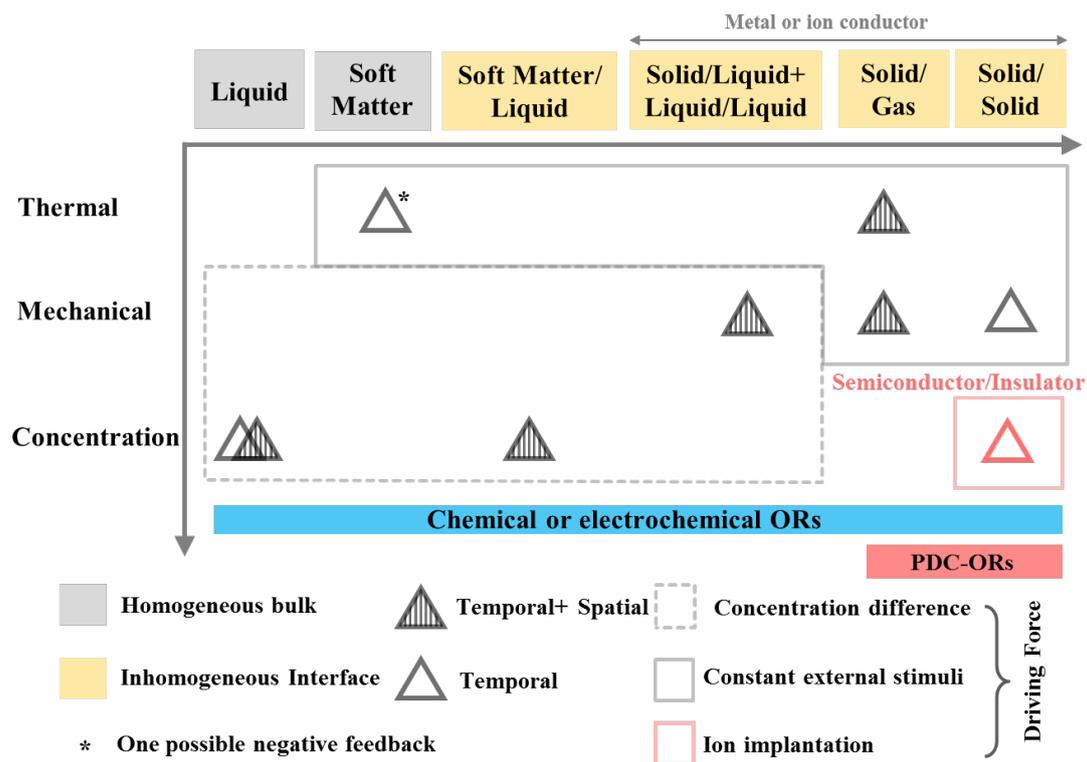

**Fig. 1| Review of ORs in literature. The triangular symbols represent known ORs together with the OR reported in this work.** Empty/full triangles represent whether the oscillation is in time or both in time and in space. The horizontal axis represents the systems in which ORs take place, and the vertical axis represents different negative feedback mechanisms. The involved matters have been roughly cataloged as liquid, soft matter (i.e., grafted polymers and gels), solid and gas. We group the ORs by driving forces with grey and red lines, and we use solid and dashed lines to indicate whether there is or there is no cross-interface matter transport in the ORs, respectively. The region inside the grey dashed line indicates ORs driven by concentration difference, the region inside the grey solid line indicates ORs driven by a constant external stimulus (e.g., electrical or temperature field) and the region inside the red solid line indicate ORs driven by ion implantation. The star symbol means that the negative feedbacks of

the marked ORs are not fully understood. Please refer to the SI for a brief summary of existing ORs.

## II Experiment

We carry out the following experiments to observe the proton-implantation induced concentration changes of the interface traps (IT) at the a-SiO$_2$/Si interface and the oxide traps (OT) inside a-SiO$_2$. These are two well-known point defects in Si-based semiconductor devices [39-43]. $N_{it}$ and $N_{ot}$ denotes the sheet concentrations, and $\Delta N_{it}$ and $\Delta N_{ot}$ their changes. Despite their extremely low concentrations (typically $10^{10}$ cm$^{-2}$ ~$10^{12}$ cm$^{-2}$ [44]), IT and OT have a large impact on the performance of semiconductor devices [45-53], making it possible to accurately obtain their concentrations using electrical measurements on the device.

We design the experiment as illustrated in Fig. 2(a). We carry out proton implantation with a customized Si-based gate-controlled lateral PNP-type transistor (GLPNP) as the sample, followed by *in situ* electrical measurement of the device to obtain $\Delta N_{it}$ and $\Delta N_{ot}$. We carry out the proton implantation and the electrical measurement in a time-staggered manner as shown in Fig. 2(b): we first ground the sample for 20 s, and then carry out the proton implantation for 1~5 μs at a constant dose rate with all electrodes grounded to eliminate the charging effect [54], after which we either connect the sample to electrical measurement instruments to perform measurement or omit the measurement step to reduce the workload. The measurement takes at most 1 min. We determine whether to take measurement by the slopes of the already measured data, and measurements are done more frequently if $\Delta N_{it}$ or $\Delta N_{ot}$ changes rapidly with respect to implantation time. This process is repeated until we reach the desired total dose of proton. We carry out the pulsed proton implantation with a plasma immersion ion implantation (PIII) system [56]. Comparing with analytic beams from accelerators [56], the proton fluence of PIII is very high, generating a high density of secondary electrons that further reduces the charging effect. Refer to the Method section for more details of PIII.

The layout and cross-section structure of the customized sample GLPNP can be seen from Fig. 2(c)-2(e). Different electrodes are marked out in Fig. 2(c) and (d), with B, C, E, G representing the base, collector, emitter and gate electrodes [57-58]. The customized GLPNP differs from regular devices by the etched-out passivation layer [bright gate region in Fig. 2(c)] and the ultra-thin electrodes with a thickness of 250 nm [Fig. 2(f)]. These changes are made to reduce the required implantation energy and to ensure a precise control of implantation depth. We set the kinetic energy of implanted protons to 80 keV according to stopping and range of ions in matter (SRIM) [59] simulations to ensure that the protons reach the a-$SiO_2$/Si interface, which is marked out in Fig. 2(e) and (f). SRIM simulation results are available in the SI.

IT and OT affect the current-voltage (I-V) relationship of the device by reducing the effective gate voltage ($V_{eff}$) through screening, allowing us to quantify their concentration changes by electrical measurement. The structure of the GLPNP is crucial for us to separate the contributions of $\Delta N_{it}$ and $\Delta N_{ot}$ from I-V curve changes due to proton implantation, where $\Delta N_{it}+\Delta N_{ot}$ is proportional to the change in the open gate voltage [$\Delta V_{G-open}$ in Fig. 2(g)] of the sub-threshold curves (STC) with a positively biased E-B junction, and $\Delta N_{ot}$ is proportional to the change in the gate voltage [$\Delta V_{G\text{-peak}}$ in Fig. 2(h)] corresponding to the current peak of the gate sweep curve (GSC) with a negatively biased E-B junction. The $\Delta N_{it}$ is then obtained from $\Delta V_{G\text{-open}} - \Delta V_{G\text{-peak}}$. Refer to the Method section and the SI for more detailed explanations.

The OT near the interface (border traps, BT) behaves differently from the OT in bulk a-SiO$_2$ due to proximity to the Si substrate [61], that OT in the bulk is always positively charged but BT can be either positively or negatively charged. The measured $\Delta N_{ot}$ contains contributions from both BT and OT in the bulk, but the concentration of BT only changes if the implanted protons are able to reach the interface. We use the notation $\Delta N_{ot+bt}$ as a synonym of $\Delta N_{ot}$ to emphasize the change of BT when appropriate in the following.

Various factors may influence the outcome of the experiment, including temperature, proton implantation parameters, reliability of electrical measurement, pristine state of the GLPNP sample, and so on. We perform a series of additional experiments to verify the validness of our results and conclusion. Details are discussed in the SI.

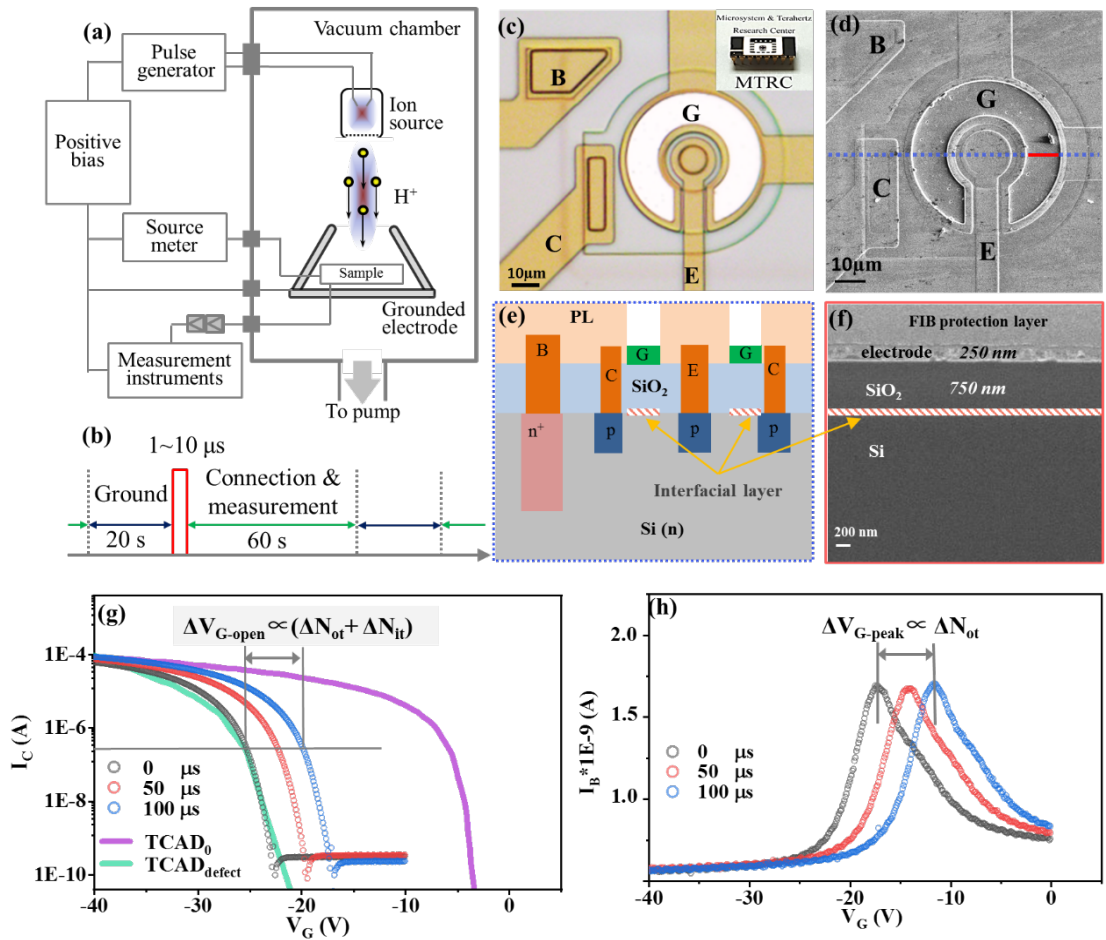

**Fig. 2| Schematics of the experiment setup, the structure of GLPNP and a demonstration of the separation of $\Delta N_{it}$ and $\Delta N_{ot}$.** (a) Schematic diagram of the combined pulsed proton implantation and *in situ* electrical measurement system. (b) Operation sequence of ground, pulsed implantation and *in situ* electrical measurement, with a rough estimate of duration; (c) Optical microscopy and (d) scanning electron microscopy (SEM) images of the planer view of the GLPNP transistor. Inset of (c) shows the package image. (e) Illustration of the cross-section of the sample GLPNP indicated by the blue dash line in (d). (f) Cross-section SEM image at the position of the red line in (d). The red/white stripes in (e) and (f) mark the interfacial region affected by implantation. (g)-(h) show the typical STCs and GSCs implanted at 80 KeV and for

0, 50, 100 µs and demonstrates how to obtain $\Delta N_{it}$ and $\Delta N_{ot}$ from the data.

## III. Results

We calculate $\Delta N_{it}$ and $\Delta N_{ot}$ from the STC and GSC measurements after each proton implantation, and we also evaluate the performance of the GLPNP by carrying out Gummel curve (GC) measurements to obtain the current gain ($\beta=I_C/I_B$, $I_B$ and $I_C$ being the base and collector currents). Both the raw data of STC, GSC and GC measurements and the processed $\Delta N_{it}$, $\Delta N_{ot+bt}$ and $\beta$ values are plotted in Fig. 3 versus the cumulative total implantation time. The specific experiment parameters are listed in the Method section.

We find that $\Delta N_{it}$, $\Delta N_{ot+bt}$ and $\beta$ in Fig. 3 all exhibit damped oscillation with a period of ~550 µs, where we estimate the period by fitting with a guessed functional form. Refer to the SI for details of the fitting. This oscillating phenomenon is unexpected and not reported in literature as far as we know. We estimate $N_{it}$ and $N_{ot}$ in the pristine sample to be $(5\pm2.5)\times10^{10}$ cm$^{-2}$ and $(2\pm0.6)\times10^{11}$ cm$^{-2}$ by matching Technology Computer Aided Design (TCAD) [60] simulations of STC with experiments [see Fig. 2(g) and the SI]. The maximum oscillation amplitudes of $\Delta N_{it}$ and $\Delta N_{ot+bt}$ are about $7.5\times10^9$ cm$^{-2}$ and $1.3\times10^{11}$ cm$^{-2}$ as seen in Fig. 3(d), (e) and the fitting results in the SI. The oscillations of $\Delta N_{it}$ and $\Delta N_{ot+bt}$ are significant since they are at the same orders of magnitude as their initial concentrations.

$\Delta N_{ot+bt}$ in Fig. 3(d) decreases in the first 250 µs for about $3\times10^{11}$ cm$^{-2}$, which is bigger than the estimated initial value of $N_{ot}$ $(2\pm0.6)\times10^{11}$ cm$^{-2}$. Aside from errors in simulation and in the matching process, this apparent discrepancy could be ascribed to the initially negatively charged BT. [61] The $N_{ot}$ obtained from the TCAD simulation corresponds to the actual value minus the concentration of negatively charged BT, leading to an underestimation.

The oscillations of $\Delta N_{it}$, $\Delta N_{ot+bt}$ and $\beta$ happen in total implantation time instead of in real time. We have checked that there is no significant change in $N_{it}$ and $N_{ot}$ during the measurement and grounding steps (refer to the SI), and they only change during the implantation. The oscillations can be seen as happening with respect to the proton dose, which is proportional to the implantation time due to the constant dose rate. We measure the effective dose using the secondary ion mass spectra (SIMS) [62] on a standard silicon slice sample accompanying the GLPNP during implantation to avoid the charging effect of SIMS, and we obtain a proton flux of $1.5 \times 10^9$ cm$^{-2}$μs$^{-1}$ or equivalently an ionization dose rate of 1.2 krad/μs. More details of the SIMS results are in the SI.

In the GC measurements, we find that $I_C$ mostly remain unchanged (see SI) and the oscillation of $\beta$ in Fig. 3(f) is entirely due to the oscillation of $I_B$. $I_B$ can be written as $I_{B-bulk} - I_{B-recovery} - I_{B-interface}$, with the three components being the bulk current, the recovery current at the E-B junction, and the interfacial current due to IT. Despite $I_B$ being dependent on these three different regions of the sample, the shape of the $\beta$ curve remains very similar to that of the upside-down $\Delta N_{it}$ curve. This shows that the oscillation of $\beta$ is entirely due to the oscillation of $N_{it}$ (the curve being upside-down since $\beta = I_C/I_B$), implying that the proton implantation induced changes mainly happen at the a-SiO$_2$/Si interface, instead of in bulk silicon or E-B junctions. We double-check this by varying the implantation depth, and we find that oscillations of $\Delta N_{it}$ and $\Delta N_{ot+bt}$ only occur when the implantation depth matches that of the interface. Refer to the SI for more details.

We summarize the observations as follows: (1) $\Delta N_{it}$ and $\Delta N_{ot+bt}$ oscillate significantly comparing with their initial values; (2) they only oscillate during proton implantation; (3) the proton implantation only affects the a-SiO$_2$/Si interface near which $N_{it}$ and some of $N_{ot}$ are located; (4) the observed oscillations are unrelated to inter-conversion between IT and OT [63] since the oscillation amplitudes differ by an order of magnitude. We therefore believe that there is enough evidence of two similar ORs of IT or OT with proton. These ORs are PDC-ORs as the oscillatory intermediates (IT and OT) are point defects in the solid and the reactions happen in the interfacial region. These PDC-ORs do not strictly follow the prediction of J. Janek [2], however, since the species that transfer across the interface are not the intrinsic point defects (IT and OT), but rather are the externally implanted protons.

Despite that IT is generally associated with dangling bond of interfacial Si atoms and OT with positively charged oxygen vacancy [43-44], their exact structures are not fully identified, so it is impossible to carry out a precise analysis of the mechanism of the two ORs directly. Nevertheless, we are able to make an educated guess of the mechanism based on experimental observations, which is presented in Sec. IV.

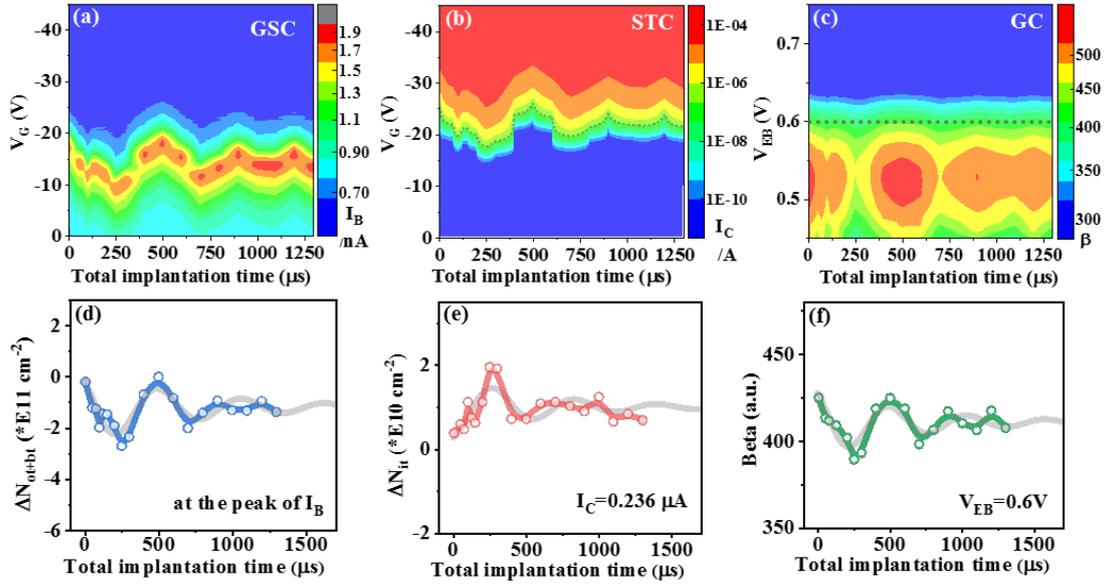

**Fig. 3| Oscillations of electric properties and of the concentration of charged point defects.** The implantation energy is 80 keV. The quantities are plotted against the cumulative total time spent during proton implantations. (a) GSC measurement results for $\Delta V_{G\text{-peak}}$, (b) STC results for $\Delta V_{G\text{-open}}$ at $I_C$=0.236 μA marked by the grey dashed line; (c) GC results for current gain β at the $V_{EB}$=0.6V marked by the grey dot line. (d) $\Delta N_{ot+bt}$, (e) $\Delta N_{it}$, (f) current gain β are obtained from (a)-(c), with the points and connecting lines representing the data, and the light grey lines representing fitting results (refer to SI).

## IV Discussion

### 1. Proton-implantation-induced reactions at the a-SiO$_2$/Si interface

Before discussing the possible mechanism of the observed oscillations of $\Delta N_{ot}$ and $\Delta N_{it}$, we first investigate the impact of proton implantation on the system. Unlike in solutions, it is much more difficult to determine the exact reactions and processes that happen inside the solid system, especially when the amorphous phase and the interface are involved. We therefore have to make educated guesses based on tests.

We need to figure out the proton-induced changes at the a-SiO$_2$/Si interface as the oscillations only happen when the implantation depth matches that of the interface. It is impractical to directly observe the changes inside the GLPNP sample *in situ*, therefore, we apply surface sensitive X-ray photoelectron spectra (XPS) and X-ray absorption spectra (XAS) on proton-implanted ultra-thin a-SiO$_2$/Si films [64-66] instead. Since the a-SiO$_2$ layers of the films are very thin (1.3 nm), XPS and XAS methods are able to yield structural and compositional information of the a-SiO$_2$/Si interface.

We carry out *ex situ* proton implantation on 4 films placed at different distances to the center of the target platform, so that the implantation doses of the films decrease as they are farther away from the center. Detailed information is shown in the Method section. We denote the film samples in increasing order of implantation dose as H-0 to H-4, where H-0 is the reference sample without implantation. The doses from H-0 to H-4 do not change linearly. The dose change from H-3 to H-4 is especially large, and the total dose of H-4 is about an order of magnitude larger than in the GLPNP experiment. The XPS and XAS curves from H-0 to H-4 are plotted in Fig. 4(a) and 4(c).

Fig. 4(a) plots the XPS peaks near the Si 2p binding energy, and the peaks change with the implantation dose. The positions of the peaks are dependent on the oxidation states of Si atoms, which are defined as the number of oxygen atoms coordinated to the Si atoms [64]. Deconvolving the XPS curves of each film yields the ratios of Si atoms with different oxidation states, including Si(0), Si(I), Si(II), Si(III), and Si(IV). The ratios are plotted in Fig. 4(b). The average composition of the interfacial layer of the pristine sample H-0 is $SiO_{1.05}$, showing that the interfacial layer is composed of sub-oxides. As the dose increases, the ratios of Si(I) and Si(II) decrease while the ratios of Si(III) and Si(IV) increase, and the average oxygen composition gradually increases to $SiO_{1.41}$ of H-4. This suggests that the proton implantation leads to remarkable oxidation of the interfacial sub-oxides, which is consistent with literatures [67-70]. We also observe a slight increase of Si(0) as the dose increases, indicating the existence of disproportionation of the interfacial sub-oxides, similar to our previous finding with Gamma-ray irradiation [64].

Fig. 4(c) plots the XAS pre-edge peaks of different films at the oxygen K edge, with reference curves of crystalline $SiO_2$ [71] and of a 50 nm a-$SiO_2$ film prepared by us. The two sites that show significant changes with respect to the implantation dose are marked with arrows. The changes mostly happen at low doses for the pre-edge peak at about 528 eV, and at high doses for the kink at about 527 eV. The XAS pre-edge peak results from the hybridization between O 2p and Si 2s, 2p [72]. This peak is very weak for the reference a-$SiO_2$ sample, while this peak of crystalline $SiO_2$ is higher than that of all the other films, indicating that the height of this peak is correlated to the local structural order of the sample. The heights of the peaks increasing with dose suggests that the structure of the a-$SiO_2$/Si interface becomes more ordered (closer to crystalline structure) after proton implantation, and a low dose is sufficient for this process to happen. This can be thought as an annealing process due to the local heating effect of incident protons [73]. The implanted protons also causes displacement damages as seen in SRIM simulations (refer to the SI), lowering the local structural order and countering the previous process, which is probably the reason why the peak does not change much from H-3 to H-4.

Implanted protons can either exist as interstitial atoms in the solid, or form bonds with Si or O atoms. The XAS kink at about 527 eV is associated with O-H bonds [74]. It only shows up in the H-4 sample indicates that the implanted protons do not bond with oxygen atoms unless the dose is much higher than that of *in situ* implantation experiments of GLPNP. The a-SiO$_2$/Si interface can be expected to be abundant with Si dangling bonds, so we expect that the Si-H bonds are more likely to form than protons being interstitials.

We also carry out density-functional theory (DFT) [75-80] calculations to verify the possible reactions from the energetics point of view. The calculated formation energies of various structures are plotted in Fig. 4(d). We calculate three groups of amorphous sub-oxide systems, where the groups differ by the percentage of hydrogen atoms, and the systems in a group differ by the percentage of oxygen atoms. The maximum percentage of hydrogen is much larger than what is possible in the film experiments. We find that oxidation and disproportionation reactions are always energetically favorable independent on the hydrogen percentage, and this agrees with the XPS tests. Fig. 4(d) also show that dehydrogenation is an energy-lowering process, suggesting that the hydrogenation process observed in XAS is only made possible by the external energy deposited by the incident protons [81-82], and the dehydrogenation process need to be considered when discussing the PDC-OR mechanism as well.

We summarize the possible reactions at the a-SiO$_2$/Si interface during proton implantation: (1) oxidation and (2) disproportionation of the interfacial sub-oxides, (3) improvement of the local structural order, displacement damage where (4) an oxygen atom or (5) a silicon atom is knocked off by proton, (6) hydrogenation/passivation of the Si dangling bond, (7) dehydrogenation/depassivation of the Si-H bond. The existence of these reactions are supported by literature as well [40-44, 56-58, 64, 83-92]. We figuratively write the reactions as:

$$SiO_{\delta-1} + O_r \rightarrow SiO_\delta, \quad (R1)$$

$$2SiO_\delta \rightarrow SiO_{\delta-1} + SiO_{\delta+1}, \quad (R2)$$

$$SiO_\delta \rightarrow {}^{LO}SiO_\delta, \quad (R3)$$

$$SiO_{\delta+1} \rightarrow SiO_\delta + O_r, \quad (R4)$$

$$Si_{\sigma+1}O_\delta \rightarrow SiO_\delta + \sigma Si_r, \quad (R5)$$

$$SiO_\delta + H \rightarrow HSiO_\delta, \quad (R6)$$

$$HSiO_\delta \rightarrow SiO_\delta + H, \quad (R7)$$

where subscript r denotes energetic recoil atoms, and superscript LO denotes locally ordered structure. $\delta<2$ since the interface consists sub-oxides. $SiO_\delta$, $SiO_{\delta-1}$, $SiO_{\delta+1}$ and $Si_{\sigma+1}O_\delta$ all represent the interfacial sub-oxides, and the different notations is merely for the convenience of writing the reaction equations.

We must stress that R1~R7 are very crude representations of the reactions since they do not reflect the actual structures of the interface. The sub-oxides ($SiO_\delta$, $SiO_{\delta-1}$, $SiO_{\delta+1}$, $Si_{\sigma+1}O_\delta$) in R1~R7 all contain active sites, and at least IT is included in these sites, so these reactions can be seen as reactions of IT. In the following, we discuss the possible mechanism for the IT oscillation, and the discussion for OT (especially BT) is available in the SI.

## 2. The possible mechanism of the $N_{it}$ oscillation

Because of the mobilities of species in solid systems are in general much lower than those in solution systems, reactions in solid systems can be difficult to take place if they involve more than one reactant. The oscillating species of the PDC-ORs in this work are immobile point defects (IT and OT), which are unlikely to take part in chemical reactions in normal conditions. We check this by carrying out GSC and STC measurements repeatedly on the GLPNP samples right after proton implantation and compare the results (see the SI). These measurements take about 1 hour in total and the measured voltages have very small variation during this time, showing that the aforementioned reactions cannot happen without proton implantation.

Besides the rate constant, the rate equation of a reaction in solution only depends on the concentrations of reactants [e.g., Eq. (1) and (2)]. It is generally impossible to write the rate equations of solid reactions this way, as the reaction rates would be dependent on the concentrations and mobilities of the active sites, on the local conformations, and so on. We argue that this does not apply to the PDC-ORs of this work. Since the proton flux is very high in our experiments, one can view the a-$SiO_2$/Si interface and active sites such as IT as being immersed in the flow of protons during the implantation. The reactions with protons can then happen despite the active sites in the solid being largely immobile, and the reaction rates would be determined by the concentrations alone similar to solution systems.

Based on this argument, we treat R1~R7 literally as if they happen in the solution despite them being crude representations of actual reactions. We then find that the structure of the reactions resembles the Brusellator model B1~B4 with minor differences. Since IT is part of the active sites in the interfacial $SiO_\delta$ layer, we use the notation IT and $SiO_\delta$ interchangably in the following. The intermediates X and Y of the Brusselator are the IT ($SiO_\delta$) and passivated IT (H-IT or $HSiO_\delta$). Reaction B1 represents the generation of the intermediate X from reactant A, and it can match the displacement damaging processes R4 and R5, considering that IT is generally thought as dangling bonds on Si atoms. Reaction B3 represents the conversion between intermediates X and Y, which matches the hydrogenation reaction R6. Reaction B4 represents the generation of the final product E, which matches the improvement of the local structure R3 since the crystalline structure does not contain $N_{it}$.

There is no direct match for B2. However, we notice that the sum of R1, R2, R4 and R7 is

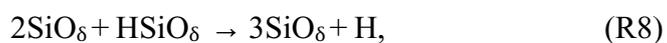

$$2SiO_\delta + HSiO_\delta \rightarrow 3SiO_\delta + H, \qquad (R8)$$

which matches the form of B2. While normally the rate equation cannot be written according to a non-elementary reaction equation, R8 might still be a reasonable match to B2, considering that the trimolecular step B2 can be explained as an approximation to multiple elementary reactions [19].

A Brusselator-like model can therefore be considered as a possible explanation of the PDC-OR of IT:

A(sub-oxides) → X(IT) + by-products     (B1')

2X(IT) + Y(H-IT) → 3X(IT) + H,     (B2')

B(H) + X(IT) → Y(H-IT),     (B3')

X(IT) → E($^{LO}SiO_\delta$).     (B4')

As explained before, B1' can be either R4 or R5. The rate equations for $N_{it}$ and [H-IT] are

$$\frac{dN_{it}}{dt} = k_B1'[A] + k_B2'N_{it}^2[H-IT] - k_B3'N_{it}[H] - k_B4'N_{it},$$     (3)

     (4)

where k denotes rate constants. [A] and [H] can be treated as constants as they (sub-oxides and implanted protons) exists in large quantities. Eq. (3) and (4) can then be solved numerically. Fig. 4(f) plots one solution, showing that this Brusellator-like model can indeed lead to oscillations with certain sets of parameters.

We stress again that the Brusellator-like model presented here is merely an educated guess of the mechanism. We are unable to further check the validity of the model since the rate constants of R1~R7 cannot be obtained with available experimental techniques, and more processes and reactions may exist. It is entirely possible that the mechanism of the PDC-OR turns out to be different from the Brusselator-like model with further investigations. Nevertheless, the Brusellator-like model is the first attempt at explaining the observed PDC-OR qualitatively, and it is consistent with all experimental and simulation evidences known to us.

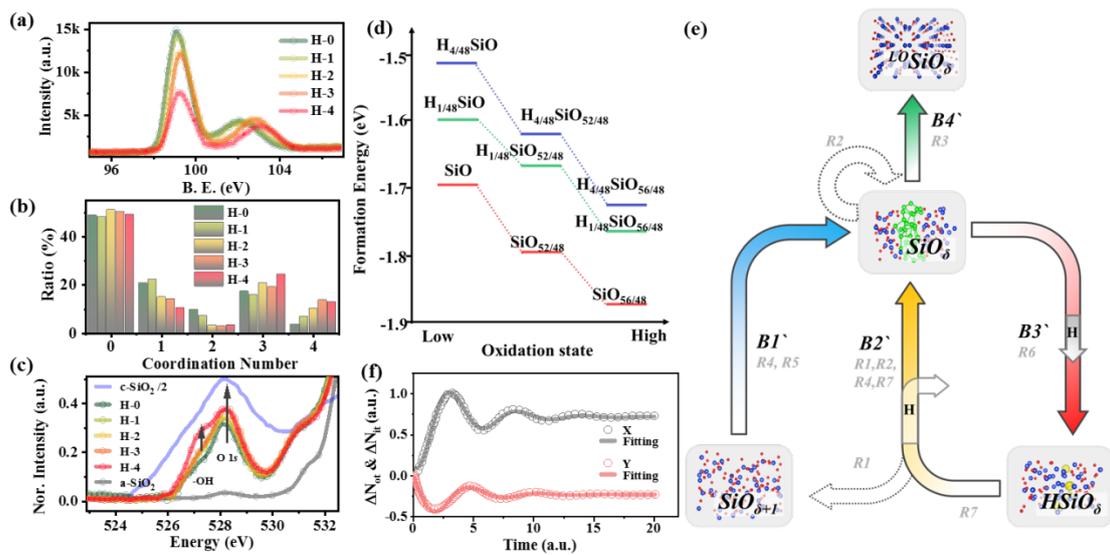

**Fig. 4| Analysis of proton implantation induced PDC-ORs.** (a) XPS spectra near the Si 2p binding energy of ultra-thin a-SiO$_2$/Si films implanted at increased proton doses from 0 of H-0 to the highest dose of H-4. (b) Ratios of different Si oxidation states of different implanted samples. (c) Pre-edge peaks of O K edge XAS spectra, together with that of crystalline SiO$_2$ (the intensity is divided by 2) [74] and that of a 50 nm a-SiO$_2$ film. (d) DFT formation energies of model systems of interfacial sub-oxides. (e) Illustration of the Brusellator-like model of PDC-OR. (f) A specific numerical solution (circles) of the Brusellator-like model, shown together with fitting results (solid lines) using the fitting function in the SI. Initial [X] and [Y] in the model are set to 1. The [A] in the model is fixed at 0.9, and the reaction coefficients $K_{B1}$~$K_{B4'}$ are set to 1, 1, 1 and 1.5, respectively. All quantities are in arbitrary units.

## V Conclusion

We present experimental evidences of a new solid-based OR in this paper. The measurement of ultra-low concentrations inside the solid is made possible using techniques common in the electrical characterization of semiconductor devices. The new OR is particularly interesting since many of its aspects are unusual and reported for the first time as an OR. It happens at the insulator/semiconductor interface, where the mobilities of species are very low. The oscillating species are point defects, making the reaction the first realization of the PDC-OR predicted by H. Schmalzried et al. [1] and J. Janek et al.[2] The oscillations are driven by the proton implantation, which provides both extra energy for endothermic reactions and a reactant in the OR.

We verify through experiments that the OR only happens during proton implantation and when the implantation depth matches that of the a-$SiO_2$/Si interface. We argue that the OR can be treated similarly as ORs in solutions due to the presence of ultra-high fluence proton flow, and we propose a simple Brusellator-like model as the first attempt at explaining the mechanism of the OR. The model agrees with all our experimental and simulation results, and it is able to reproduce the observed damped oscillation behavior. The model is better suited to explain the oscillation of $N_{it}$. We also apply the model to $N_{ot}$ in the SI, but there is no suitable way to verify the validness of the model for $N_{ot}$. We are going to further investigate the mechanisms of the PDC-ORs presented in this work in the future with newly developed experimental techniques for solids.

Ion implantation is a common technique employed in the fabrication of semiconductor devices, but its use in physics and chemistry studies is not much in the past. We demonstrate in this work that ion implantation can lead to new phenomena that has never been reported in literature, so it might be worthwhile to further investigate its use in fundamental researches. Aside from that, there have been a lot of method studies trying to introduce proton into systems (such as ionic liquid gating [93-96], hydrogen decomposition catalyzed by noble metals [97], and non-catalytic acidic solution soak [98]), since proton plays an important role in the property control of functional materials [82, 98] and devices [83, 93-94, 99], and we believe that proton implantation can be an alternative method for these applications as well.

## VI Method

### 1. Fabrication of the GLPNP and ultra-thin a-SiO2/Si films.

The GLPNP is fabricated at the National Laboratory of Analog Integrated Circuits (China Electronics Technology Group Corporation, Chongqing, China) following a standard bipolar linear integrated circuit process flow. The passivation layers (1 μm $Si_3N_4$ and 1 μm $SiO_2$) right above gate oxide ($SiO_2$, ~750 nm) is etched off. The thickness of metal electrodes are controlled to be around 250 nm, which is much smaller than the ~1.5 μm of common transistors.

Ultra-thin a-$SiO_2$/Si films of 1.3 nm were obtained by applying reactive ion etching [A63] on thick $SiO_2$ thick films, which are fabricated with the same process parameters to that of gate oxide in the GLPNP.

All the thickness values are verified using a GAIA3 FIB-SEM dual beam system after fabrication of the cross-section.

### 2. *In situ* pulsed and *ex situ* proton implantation

The *in situ* implantation system consists of a high voltage pulsed power generator, a PIII chamber and an electrical measurement sub-system. The proton energy can range from 10 to 150 KeV with the independent pulse width of 1~5 μs [100]. The actual fluence range for the 80 KeV experiments is 1.2~1.9×$10^{10}$ $cm^{-2}$ $μs^{-1}$, which is measured by SIMS on deuterium implanted Si standard sample. A positive bias between the ion source and GLPNP is applied to drive the proton to the GLPNP target. Both the distance and the bias voltage between the ion source and the GLPNP are kept constant as ~8 cm and 100 V if not specified otherwise.

We carried out *ex situ* proton implantations on ultra-thin a-SiO$_2$/Si films at an energy of 110 eV and a total fluence of $4.5\times10^{14}$ cm$^{-2}$. The atmosphere pressure in the chamber is around $10^{-5}$ Torr. The proton plasma of PIII is mainly concentrated in a circular region with a radius of 10 mm, and the proton fluence decreases gradually with the increasing distance to the center of the circular region. Films H-4 to H-1 with a size of 5×5×0.5 mm are placed on the target platform with the distance to the center being 0, 15, 30, and 45 mm, so that the samples are implanted with decreasing doses of protons.

### 3. Electrical and Electronic Measurements

All electrical measurements (STC, GSC, GC) are carried out in a portable high precision electrical measurement system (Platform Design Automation, Inc., Beijing) with four independent source measurement units (National Instrument). We collect all the voltage-dependent currents (E, B, C and G) during measurements. Since the currents can be extremely small (pA level), we use low resistance coaxial cables and carefully ground the system to guarantee accurate measurement.

XPS measurements are performed in an ESCALAB 250 system using Al target. The vacuum is maintained at $1\times10^{-7}$ Pa. All the samples are in good electrical contact with the sample holder and an electron flood gun is employed to neutralize the surficial charges. The binding energy has been calibrated using the C 1s peak as standard. XAS is collected in the BL12B-a Station (National Synchrotron Radiation Laboratory, China) at the total electron yield mode. The effective detection thickness is less than 5 nm. Five 1.3 nm a-SiO2/Si films named as H-0~H-4 and one 50 nm a-SiO2/Si film are measured at the same batch. The X ray beams of both XPS and XAS are about 0.5*0.5 mm, and are calibrated to point to the center of the sample before measurement.

### 5. Theoretical Calculation and Simulations

The *ab initio* formation energies are calculated using pseudopotentials [77] with the HSE06 functional [78] as implemented in the PWmat software [75-76]. We use homogeneous supercells with 48 Si atoms to represent $SiO_{1+x}$ (x=0, 4/48, and 8/48). The $SiO_{1+x}$ supercells are constructed by fully relaxing the structure obtained from removing oxygen atoms randomly from a previously optimized a-$SiO_2$ supercell. We then add hydrogen atoms to construct fully relaxed $H_ySiO_{1+x}$ cells (y=0, 1/48 and 4/48).

The TCAD simulations of the GLPNP are carried out using the Silvaco [60] software on a model (see SI) with the same structure and dimensions as the actual device [101]. We match the simulated GSC and GC results with experiments by adjusting $N_{it}$ and $N_{ot}$. SRIM simulations are carried out on both regular and customized GLPNP to determine the proper energy of interfacial implantation. See the SI for more detailed information of TCAD and SRIM simulations.

## Data availability

The data shown in the paper are available from the corresponding authors on reasonable request.

## Acknowledgement

This work was supported by the National Natural Science Foundation of China (11805169, 11804314, 11991060, and 12088101), NSAF Program of China (U1930402) and Open Program of State Key Laboratory of Nuclear Physics and Technology, PKU (NPT2020KFJ17). D. Meng appreciates the kind help from Dr. Lechen Chen of Sichuan University. This work was partially carried out at the USTC Center for Micro- and Nano-scale Research and Fabrication.

## Author contributions

D. M. designed the research. D. M., G. Z., and M. Li both developed the methodology and carried out the *in situ* measurements. D. M., and G. Z. analyzed the results of electrical measurements. H. Z, and D. M. designed the GLPNP sample. D. M., and M. Li stabilized and analyzed the ion beam. D. M., and Z.-H.Yang conducted the literature search. S. H. L. C., and D. M. collected and analyzed the XPS data. H. D., and W. Y. performed the XAS measurements . M. Lan carried out DFT calculations. C. J. performed the TCAD simulations. G. Z., and H. Z. carried out *ex situ* temperature dependent measurements. D. M analyzed the SIMS, XAS and XPS data. D. M., and X. Z. constructed the Brusellator-like model. Y. S., and Y. L. contributed to the development of the fitting function and Brusellator-like model. G. D. and Y. D. provided administrative, technical and material support. D. M., Z. -H. Y., G. D., and S. -H. W., wrote the manuscript with contributions from J. X., and L.C. All authors are involved in the discussion of the data and the conclusions. All authors contributed to the improvement of the manuscript, and have approved its final version.

## Corresponding authors

Correspondence should be addressed to D. M., Z.-H. Y., G. D. or S.-H. W.

**Competing interests.**

# Supplementary Information for

Emerging Oscillating Reactions at the Insulator/Semiconductor Solid/Solid Interface via Proton Implantation

## 1. A brief review of ORs in Figure 1 of the main text

In Fig. 1 of the main text, we classify the ORs in literature according to various aspects of these reactions. In this section, we provide references and a brief review of all the ORs in Fig. 1. We present the ORs from the left side to the right side of Fig. 1. In this paper, we roughly categorize all the material states into liquid, soft matter (i.e. polymer, gel and liquid crystal), solid and gas based on the existence forms.

1.1 ORs in liquid

ORs in liquid (aqueous solutions) is the most common and typical type of ORs, and their mechanism is the well-studied among all ORs. The most well-known ORs in solutions include the Belousov–Zhabotinsky reaction (BZ) [1-3], the Bray–Liebhafsky reaction (BL) [4], the Briggs–Rauscher reaction (BR) [5] and Bromate–Sulfite–Ferrocyanide reaction (BSF) [6]. Some of these ORs have intermediates with visible color, and one can observe the periodic color changes when the solution is stirred and homogeneous. If the solution is not stirred, the diffusion of intermediates would turn the temporal oscillations to spatial ones and chaotic patterns would develop, and therefore we label these ORs in Fig. 1 as both oscillating in time and oscillating in time and space. There is no phase transition (i.e. liquid to solid) for both cases, so we mark these ORs in Fig. 1 as reactions happening in homogenous liquid.

Due to the high mobility of the species in aqueous solutions, these ORs do not need an external stimuli to sustain the far-from-equilibrium state. The concentration differences of reactants and intermediates not only drive the ion diffusion, but also serve as negative feedbacks to driving the reaction away from the equilibrium.

In addition to these inorganic reactions, organic ORs could also happen in liquid. Sergey N. Semenov et al. [7] reported an organic ORs in aqueous solutions, which shows auto-oscillations without external stimuli. Its mechanism involves a series of reactions similar to the inorganic ORs.

H. Sugiura et al [8] reported a special case of BSF OR, which happens in a micro-fluid reactor. Only under the electrical filed, the chemical reactants separated by oil could contact with each other and make the OR happen. Since the electrical filed only plays the role of mixing the reactants instead of providing energy to make the reaction happen, we still label it as an OR in liquid without external stimuli.

1.2 ORs in soft matter

Kamlesh Kumar et al. [9] reported a chaotic oscillation that a constant external light stimuli (e.g. sunlight) cause a continuous motion of a polymer belt (in solid states), which consists of liquid crystalline networks (LCNs) and photosensitive fluorinated azobenzene molecules (F-azo). The LCNs and F-azo are almost homogeneously mixed and then polymerized together. Thus the polymer belt could be taken as a homogenous solid bulk. Although the mechanism of the ORs is still not fully understood, Kamlesh Kumar et al. guessed that the local variation of temperature might contribute to the structural changes of F-azo. Thus we preliminary put the symbol of negative feedback on the thermal region in the Fig. 1. We therefore catalog this oscillating phenomena to homogenous solid based ORs under constant external stimuli.1.3 ORs at the soft matter/liquid interface

Chemomechanical oscillator, firstly reposted by R. Yoshida [10], is one of the common example of ORs happens at the interface between the liquid and soft matter (polymers, gels or liquid crystal) [11-14]. The catalyst ions (e.g. Ru ions) of BZ reaction could be grafted to or embedded in the soft matters. When the BZ reaction mixture solution is in contact with the soft matter, the ORs would happen at the soft matter/liquid interfaces. Moreover, if the soft matter are sensitive to the external stimuli, such as $H^+$, the oscillating concentration of the intermediate (H+) of the BZ OR would cause oscillating shape changes of the polymers or gels, which introduces an interesting application nicknamed as "atomic brushes".

Consistent with the the BZ OR in liquid, the solid/liquid interfacial BZ ORs also have the concentration difference as the negative feedback and do not need constant external stimuli (driving force) to make the OR happen.

1.4 ORs involving both Solid/liquid and liquid/liquid interfaces

An oscillating phenomenon nicknamed as "beating heart", involves a liquid metal drop (such as mercury [15] or gallium [16]), an acid solution and a solid metal (such as iron) that barely touches the drop. The shape of the liquid metal drop oscillates in time due to a coupled mechanism involving both chemical reactions and mechanical feedback. The reactions happen at the liquid/liquid and liquid/solid interfaces individually, and they are coupled by the shape changes of the liquid metal drop. We therefore represent this reaction as "solid/liquid+liquid/liquid ORs" in Fig. 1.

1.5 ORs at the solid/gas interface

M. Lallemant et al. [17] observed oscillations in the high temperature oxidation of solid Ti-Zr alloy. External stimuli (high temperature) is required to increase the mobility of solid atoms or ions and to make the reaction happen. The mechanism is not fully understood yet. There are different mechanisms to interpret the oscillating phenomena based on the thermal or mechanical feedbacks as described in the following.

Within a proper temperature window, the surface temperature of Ti-Zr alloy oscillates, together with the formation of multilayered products consisting of two different phases. The oxidation (e.g. forming $TiZr_2O_4$) is not only endothermic, but also change the stoichiometry (more Ti but fewer Zr) of the interface between the oxidized layer and the metal. The decreased temperature together with higher activation energy due to changed interface composition slow down the oxidation reaction. We classify this as the thermal feedback. The oscillating phenomenon can be interpreted with a mechanism involving mechanical feedback as well. The accumulation of the stress at the interface between the oxidized layer and metal prevents the transport of the metal atoms and hinder the oxidation reaction. Once the stress relaxes, the oxidation would happen again. Either mechanism would cause the formation of multilayer at the surfaces.

The mechanical feedback mechanism is also supported by the high temperature corrosion of Ni reported by Pierraggi [18] and Ti by G. Bertrand [19]. From the atomic point of view, the diffusion of metal atoms is not negligible at the interface between the oxidized layer and the metal, which would lead to vacancies in the metal at the interface. The continuous formation of vacancies leads to interfacial voids, which hinder the further diffusion of the metal atoms. The interfacial voids would collapse due to stress build-up and lead to oscillating behavior.

1.6 Solid/solid interface.

The oscillating phenomena of electrical currents in chemical cells are known for a long time. As summarized by J. Janek [20], these ORs could happen at the AX/BX (both are cation conductors), AY/BY (both are anion conductos), (A,B)/AX (interface between metal/alloy and ion conductors) and even the interface between AX and complex oxide (e.g. YSZ). They involve electrolysis and electroplating at the interfaces between two solid ion conductors or between metal and solid ion conductors.

In these processes, the formation of interfacial void agglomerating from point defects plays the role of mechanical feedback that hinders the subsequent ion transferring across the interface, and the collapse of the void would rebuild the connective interfaces and allow the ion transfer again. Such a feedback mechanism lead to an oscillating overvoltage of the metal electrode [20]. We therefore represent these reactions as solid/solid interfacial ORs driven by the external stimuli (electrical field) in Fig. 1.

## 2. SRIM Simulations

The passivation layer of a regular GLPNP transistor contains two sub-layers (1 μm $Si_3N_4$ and 1 μm $SiO_2$), and the thickness of the top metal electrode is about 1.5 μm. A much larger implantation energy than that in the main text would be needed for protons to reach the interfacial region of a regular GLPNP. Limited by the energy range of our PIII system, we customize the GLPNP by etching-out the passivation layer and design a much thinner top electrode (250 nm) to reduce the needed implantation energy. To avoid the corrosion of the aluminum alloy metal electrode, the GLPNP chips are preserved in a nitrogen cabinet and the packaged ones in a drying cabinet with constant temperature. We find that the electrical performances of the GLPNP have no visible change even after one year storage. The electrical data of the GLPNP in this manuscript are collected within half a year after fabrication. We simulate the implantation-energy-dependent proton distribution in regular and customized GLPNP using SRIM software, and the results are shown in Fig. S1(a) and (b), respectively. By etching out the passivation layer and using a thinner top electrode, the distance traversed by the protons to reach the a-$SiO_2$/Si interface is shortened from 4.25 μm to 1 μm, leading to a vast reduction of implantation energy from 350 keV to 80 keV. Moreover, the lower implantation energy results in a higher doping ratio and a narrower ion distribution peak.

Energetic ion implantation into a solid would generate numerous defects both on the surface and in the bulk. Since the top surface of the customized GLPNP is metal electrode, the surface damages, such as sputtering and roughening, have few influence on the electrical characteristics of GLPNP. In the bulk, ion implantation can lead to the generation of vacancies, interstitials and voids. Fig. S1(c) shows the simulated formation of Si and O atomic vacancies when the implantation energy is 80 keV. Layer 1~3 in Fig. S1(c) denotes the Al alloy electrode (250 nm), the a-$SiO_2$ layer, and the Si substrate respectively. A small part of the O vacancies that are electrically active $N_{ot}$ in electrical measurements. The implantation would also induce diffusion and reaction (such as recombination) of these defects. The O atoms knocked out by the implantation process is the oxygen source of interfacial oxidation (marked as reaction "R1" in main text). Compared with the damage in bulk a-SiO$_2$, much fewer Si recoil atoms are generated in the Si substrate, suggesting that the displacement damage in Si is negligible when the implantation depth matches the interfacial region.

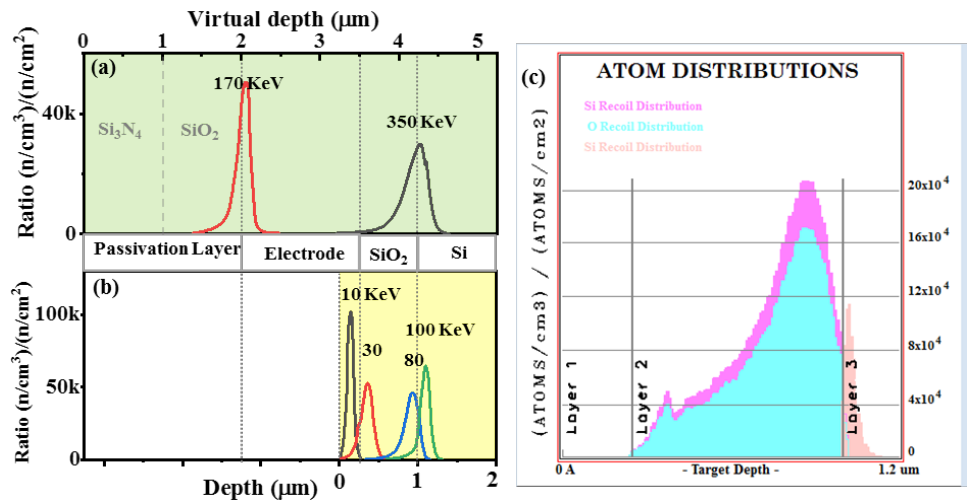

**Fig. S1|** **SRIM simulations of the thickness depth of proton and atomic vacancies.** The implantation energy dependent proton distribution on (a) regular GLPNP with 2 μm passivation layer and thick top electrode and (b) our customized GLPNP. (c) SRIM simulation of the distribution of implantation-induced atom recoil in a-SiO$_2$ (layer 2) and Si (layer 3) with an implantation energy of 80 KeV.

## 3. Quantification of $N_{it}$ and $N_{ot}$

The identification and quantification of $N_{it}$ and $N_{ot}$ are based on the classical theory of semiconductor physics and are well accepted in the characterization of electronic devices [21].

The GLPNP can be viewed as a hybrid of a bipolar junction transistor (BJT) and a metal-oxide-semiconductor field-effect transistor (MOSFET). The emitter (E), the base (B) and the collector (C) parts of the GLPNP constitutes a BJT. The gate (G), C and E parts constitutes a MOSFET with the C acting as the source (S) and E as the drain (D). By changing the interfacial band bending, the gate voltage ($V_G$) controls the working state (on or off) of the MOSFET and change the base current together with the current gain ($\beta=I_C/I_B$). The effective voltage $V_{eff}$ would equal to $V_G$ if there is none IT and OT in an ideal system. In the real system, charged OT and IT would screen the gate voltages as shown in Fig. S2(a).

Fig. S2 (c) illustrates how the STC measurement works. The band bending near the a-SiO2/Si interface can be changed by applying gate voltage, including the energy level of the intermediate (middle) level and the Fermi level. Fig. S2(c) also marks out the surface potential level $\varphi_S$, which is an intrinsic property determined by the size of the transistor. We denote the difference between $\varphi_S$ and $E_i$ by $V_{G1}$. $V_{G1}=0$ means that the transistor is in the open state, and we can calculate the theoretical collector current $I_{c\text{-open}}$ to be 0.235 µA [22]. According to the value of $I_{c\text{-open}}$, we could obtain the real open gate voltage $V_{G\text{-open}}$ in the sub-threshold curve (STC) measurement ($V_{EB}$=0.5V, $V_{CB}$=0), with Fig. 2(g) of the main text as an example. The changes of open voltage $\Delta V_{G\text{-open}}$ could be ascribed to the changes of the concentrations of charged point defects $\Delta N_{it} + \Delta N_{ot}$.

Fig. S2(b) and S2(d) describe the working of the GSC measurement. We denote the difference between $E_i$ and $E_f$ by $V_{G2}$. The $N_{it}$ is not charged when $E_i$ and $E_f$ are aligned with $V_{G2}=0$. As described in the main text, both the $I_{B\text{-interface}}$ and $V_{Nit}$ could be ignored. It results in a peak of $I_B$ in gate sweep curve (GSC) measurements ($V_{EB}=-1V$, $V_{CB}=0$) with the example measurement result shown in Fig. 2(h). The gate voltage at the $I_B$ peak is denoted as $V_{G\text{-peak}}$, whose changes are only caused by $\Delta N_{ot}$. The conversion factor between the changes of gate voltage and that of charged defect concentration is determined by the size of the GLPNP, which is 3.8E10 cm$^{-2}$V$^{-1}$ for this work. We then obtain $\Delta N_{ot}$ and $\Delta N_{it}$ separately by combining GSC and STC measurement results. The current gain factor β (typically at $V_{EB}=0.6V$) of BJTs could be obtained in the measurement of the Gummel curve (GC) ($V_{G\text{-}B} = V_{C\text{-}B} = 0V$) as shown in Fig. 3(g).

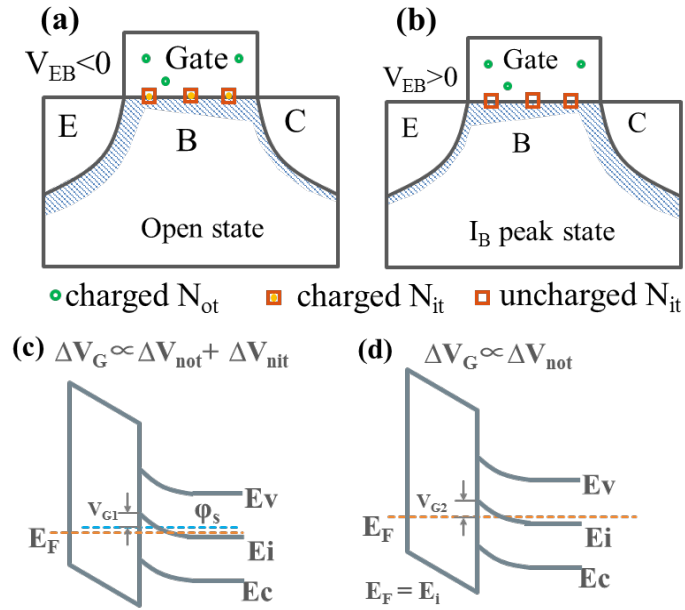

**Fig. S2| Illustration of the STC and GSC measurements and the separation of ΔNit and ΔNot.**

(a) and (b) show the electronic states of STC and GSC respectively, corresponding to different bias states of E-B junction and different charging stats of $N_{it}$. Interfacial band bending are plotted for (c) STC and (d) GSC. $E_V$, $E_C$, $E_i$, $E_F$ and $\varphi_S$ represent the valence band energy, conduct band energy, intermediate energy, Femi energy, and the surface potential.

We use TCAD simulations to determine the initial $N_{it}$ and $N_{ot}$ of the GLPNP. TCAD is a physics-based device modeling technique that is able to simulate the performances of electrical devices over a range of conditions. A three dimensional model as shown in Fig. S3(a) is constructed based on the dimension, material type, contacts, and doping concentration of the actual GLPNP. We use SEM and SIMS measurements to verify the dimension and the doping concentrations of the model. Fig. S3(b) shows the cross-sectional view of the TCAD GLPNP structures. Simulation is performed using the finite element method [23]. We match the simulation with experimental STC [Fig. 2(g) of the main text] and GC [Fig. S3(c)] results by adjusting the input parameters including bias conditions, material parameters and $N_{it}$ and $N_{ot}$ concentrations.

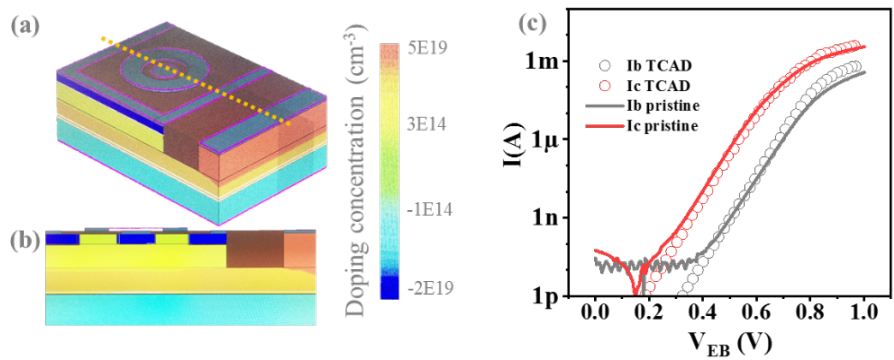

**Fig. S3| Model and data matching of TCAD simulations.** (a) The 3D model of the GLPNP with dimensions and parameters double checked by the SEM and SIMS measurement. (b) The cross-section of the GLPNP model along the yellow dashed line of (a); (c) A comparison of the simulated and measured I-V curves of GC measurements.

## 4. Fitting with the damping oscillation function

Although the Brusellator model do not have a closed-form solution, the damped oscillation seen in Figs. 3(d)-(f) in the main text and Fig. S4 can be roughly described by a simple fitting function as $y(F) = p + A\exp(-F/F_h)\sin(\omega F + \varphi)$. Fitting the data yields the approximated periods and amplitudes of the oscillations, and the fitting parameters also allows simple comparison between different data. The fitting parameters of the data in Fig. 3 and Fig. S4 are shown in Table S1, and those of Fig. 4(f) of the main text are shown in Table S2. Fig. S4 demonstrates the fitting of IB and β of Fig. 3 of the main text. Fig. S4 also plots $I_C$ and shows that it does not change significantly during the implantation and β results from the oscillation in $I_B$. The result further proves that the implantation causes few damage of the Si bulk.

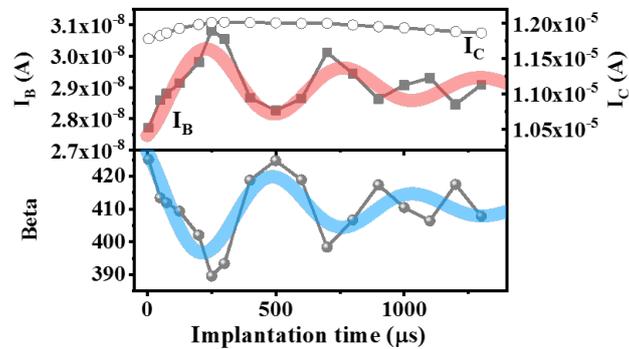

**Fig. S4| Demonstration of the fitting function using $I_B$ and β**. The data is the same as in Fig. 3 of the main text. The colored lines are the fitting lines, and the points are the experimental data. We plot

the $I_C$ data as well to show that its fluctuation is small comparing with its initial value.

Table S1 Fitting parameters of $I_B$, $β$, $ΔN_{it}$ and $ΔN_{ot+bt}$ of Fig. 3 of the main text.

|  | *p* | *A* | $F_h$ | *(2π/ω)* | *φ/π* |
|---|---|---|---|---|---|
| $I_B$ | (2.9±0.11)E-8 | (1.6±0.35)E-9 | 712±249.5 | 546±30 | 0.3±0.17 |
| $β$ | 410.3±0.26 | 20.3±0.47 | 738±263.2 | 548±30 | 1.3±0.16 |
| $ΔN_{it}$ | (9.9±0.22)E9 | (7.5±0.52)E9 | 717±338.2 | 556±44 | 0.3±0.25 |
| $ΔN_{ot+bt}$ | (-1.3±0.13)E11 | (1.3±0.42)E11 | 733±227.3 | 566±28 | 1.3±0.14 |

Table S2 Fitting parameters of the numerical solution of the Brusselator-like model of Fig. 4(f) of the main text.

|  | *p* | *A* | $F_h$ | *2π/ω* | *φ/π* |
|---|---|---|---|---|---|
| X | -0.24±0.002 | 0.37±0.013 | 3.38±0.167 | 5.68±0.030 | -0.23±0.072 |
| Y | 0.71±0.003 | 0.70±0.014 | 3.75±0.136 | 5.56±0.030 | 0.67±0.085 |

## 5. Verification for the Mechanism of the PDC-OR

As mentioned in the main text, various factors may affect the outcome of the proton implantation experiments. We carry out a series of additional experiments to verify the conclusions of the main text, and the results are presented here.

### 5.1 Implantation depth and proton fluence

We vary the implantation depth to check whether the PDC-OR only happens when the implantation depth matches the depth of the a-SiO$_2$/Si interface. We first carry out SRIM simulations to determine the implantation depth distribution corresponding to each implantation energy, which is discussed in Sec. 2 of the SI.

We then carry out proton implantation experiments with the implantation energies being 10, 30, 80 and 100 keV. According to Fig. S1(a) and (b), these implantation depths match the top electrode, the SiO$_2$ layer, the a-SiO$_2$/Si interface, and the Si substrate, respectively. The implantation dependent changes of $\Delta N_{ot}$, $\Delta N_{it}$ and $\beta$ of GLPNP implanted at 30 and 100 keV are shown in Fig. S5 (a) and (b).

The protons cannot reach the interface at 30 keV, and Fig. S5 (a) shows that $\Delta N_{it}$ is close to 0 after an implantation time of 250 µs and $\Delta N_{ot}$ changes a lot with implantation time. This is expected since $N_{it}$ only exists in the interfacial region while $N_{ot}$ exists in both the a-SiO$_2$ layer and the interfacial region. $\Delta N_{ot}$ and the current gain $\beta$ change monotonically during the implantation. When the implantation energy increases to 100 keV, $\Delta N_{it}$, $\Delta N_{ot}$, and $\beta$ all change significantly, but the changes remain monotonic even with a large proton dose corresponding to 500 µs. Although these experiments are not extensive, they suggest that only the interfacial implantation results in the oscillations of $N_{it}$ and $N_{ot}$.

The proton fluence of the *in situ* PIII experiments need to be calibrated by experiment. The fluence is determined by SIMS on a deuterium (D) implanted silicon slice using the same parameters as those of the proton implantation experiment on GLPNP as shown in Fig. S5(c). Due to the low abundance of D in nature, the sensitivity of D in SIMS can be much higher than that of H. The lower charging effect of Si than that of SiO$_2$ helps to further improve the sensitivity. The detection limit of D in silicon determined by SIMS is as low as $10^{14}$ cm$^{-3}$, which is three orders of magnitudes better than that of H ($10^{17}$ cm$^{-3}$). The measured fluence of $1.5\times10^9$ cm$^{-2}$µs$^{-1}$ is much larger than the conventional ion implantation system (~$10^2$ cm$^{-2}$µs$^{-1}$) [24].

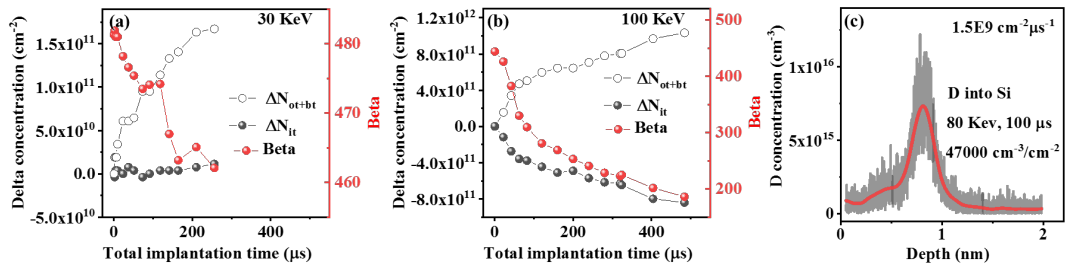

**Fig. S5|** Results of proton implantations with different implantation depth and fluence. Implantation-time-dependent $\Delta N_{ot}$, $\Delta N_{it}$ and $\beta$ of GLPNP implanted at (a) 30 KeV and (b) 100 KeV. (c) The SIMS results of D in Si, with the same implantation parameters as those of Fig. 3 of the main text.

**5.2 Reliability of electrical measurement**

We check the reliability of the electrical measurements by making repetitive GSC and STC measurements on pristine and proton-implanted GLPNP samples. The purpose of these tests is (1) to determine the magnitude of the fluctuation of the measured values, (2) to check whether previous measurements can influence the outcome of later measurements (such as the wake-up effect in ferroelectric materials [25]), and (3) to check whether the PDC-OR only happens during the proton implantations.

The results are plotted in Fig. S6. The fluctuations of the measured quantities are very small. The results of repeated measurements are highly consistent with each other. The measurements takes about 1 hour in total, and it shows that $N_{ot}$ ($\Delta V_{G\text{-open}}$) and $N_{it}$ ($\Delta V_{G\text{-open}} - \Delta V_{G\text{-open}}$) do not change during this, which suggests that the PDC-OR cannot happen without the proton implantation.

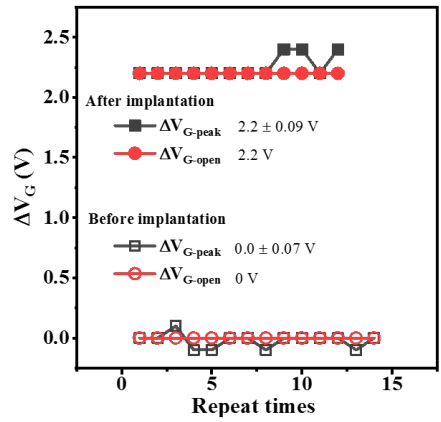

**Fig. S6| Stability of electrical performance of GLPNP in repetitive measurements.** The data for both pristine GLPNP (open squares and circles) and and implanted GLPNP (filled squares and circles) are plotted. The measurements last about one hour. The standard error of $\Delta V_{G\text{-open}}$ is 0 and that of $\Delta V_{G\text{-peak}}$ is as small as 0.07 and 0.09V respectively for pristine and implanted GLPNP.

### 5.3 Influence of the temperature

Since the performances of semiconductor devices are sensitive to the temperature, the results of the electrical measurements are liable to change due to temperature fluctuations. While our lab temperature is air-conditioned to 25°C, we still need to check the actual temperature during the electrical measurements. We carry out GSC and STC measurements of the pristine GLPNP sample at different temperatures in order to determine the temperature dependence of the electrical measurements, which in turn help us to obtain the actual temperature during the measurements of the proton implantation experiments. The results are shown in Fig. S7(a) and (b). The temperature is PID controlled using a commercial cooling box, and the temperature fluctuation is within 0.3 °C. The temperature dependent changes of the gate voltage and the peak base current are plotted in Fig. S7(c). The gate voltages and the base current all increase monotonically with the temperature. The increase of $V_{G\text{-peak}}$ of GSC and $V_{G\text{-open}}$ of STC is about 0.6 V and 0.3 V as the temperature goes from 25°C to 60°C, which is an order of magnitude smaller than the maximum amplitude of the oscillations of $V_{G\text{-peak}}$ and $V_{G\text{-open}}$ (~5V according to Fig. 3 of the main text). The measured peak base currents in the proton implantation experiment are about 1.5~2 nA as shown in Fig. 2(h) and Fig. 3(a). According to Fig. S7(c), $I_{B\text{-peak}}$ at 2 nA corresponds to 25°C and 1.5 nA to 23 °C, which means that all the measurements in the proton implantation experiment are taken at roughly the same temperature, and the observed oscillations have nothing to do with possible temperature fluctuations.

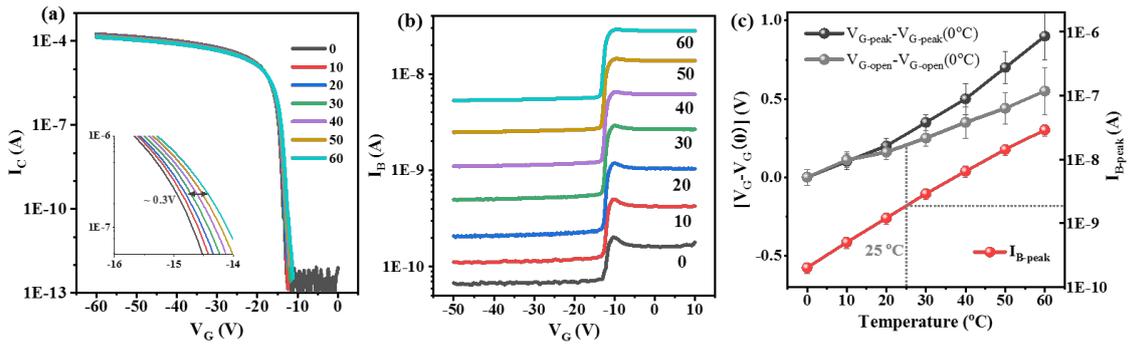

**Fig. S7| Temperature dependent electrical characteristics.** (a)-(b) exhibit the *ex situ* temperature dependent STC and GSC curves of a typical GLPNP respectively. (c) shows the temperature dependent changes of the open gate voltage and peak current.

### 5.4 Fluence and initial state dependence

We also test the influence of the proton fluence and the pristine state of the GLPNP sample. Fig. S8(a) and (b) plots the results for three additional *in situ* proton implantation experiments denoted as A, B and C.

The fluences for A and B are the same ($2.3\times10^{10}$ cm$^{-2}$µs$^{-1}$ as determined by SIMS). However, $\Delta N_{ot+bt}$ and $\Delta N_{it}$ oscillate in B but not in A. This indicates that the initial state of the sample has a strong impact on the oscillation behavior, because the GLPNP samples in A and B still differ in the local structures despite them being fabricated with the same process and showing almost similar pristine values of electronic properties. Similar initial-state dependence have been reported in the studies of radiation effects of silicon based electronic devices, which show different responses to Gamma-ray irradiation despite being fabricated in the same batch or on the same wafer [26].

The fluence for A and B is one order of magnitudes larger than that of the main text ($1.5\times10^9$ cm$^{-2}$µs$^{-1}$), and we find that the oscillations of B become irregular and are different from the damped oscillations shown in the main text. The average period of B decreases to about 6 µs ($1.38\times10^{11}$ cm$^{-2}$), which is much smaller than the ~550 µs ($8.25\times10^{11}$ cm$^{-2}$) of Fig. 3 in the main text. We effectively reduce the fluence in C by increasing the distance between the ion source and the target to 12 cm, and the period is about 10 µs. These results suggest that the period of the oscillations is inversely proportional to the fluence.

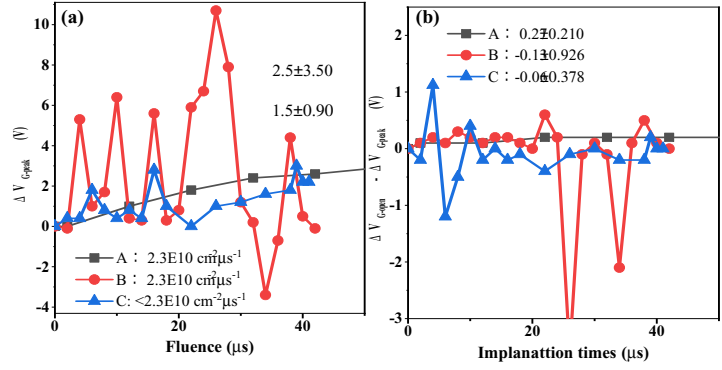

**Fig. S8|** Results of *in situ* proton implantations with larger dose rates than in the main text. (a) and (b) plot the implantation-time-dependent changes of $\Delta V_{G\text{-open}}$ and $\Delta V_{G\text{-open}}-\Delta V_{G\text{-peak}}$, corresponding to $\Delta N_{ot+bt}$ and $\Delta N_{it}$ respectively.

## 7. A Brusellator-like model for the oscillating of $N_{ot}$

We discussed the possible mechanism of the $\Delta N_{it}$ oscillation by first probing the proton-induced reactions at the a-SiO$_2$/Si interface with experiments. This cannot be done with $N_{ot+bt}$, because it is difficult to pin-point the exact locations of the reactions as OT exists in the entire a-SiO$_2$ layer. Since the shape of the $\Delta N_{ot+bt}$ curve in Fig. 3(d) of the main text is similar to the shape of the upside-down $\Delta N_{it}$ curve in Fig. 3(c), we speculate that a Brusellator-like model may be applicable to the oscillation of $\Delta N_{ot+bt}$ as well. In the following, we denote OT (including BT) as $[\frac{1}{2}V_O]H{:}[\frac{1}{2}V_O]$ based on the reactions reported by Julien Godet et al. [R23], where the two dots represent a lone pair of electrons. Julien Godet et al. [27] reported the following two reactions in a-SiO$_2$ obtained from first-principle calculations:

Reaction 1:

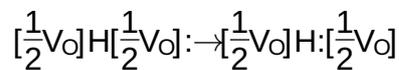

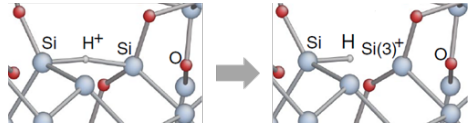

Reaction 2:

$$[\tfrac{1}{2}V_O]H:[\tfrac{1}{2}V_O] \rightarrow [\tfrac{1}{2}V_OH]:[\tfrac{1}{2}V_O]$$

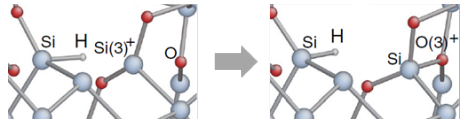

Aside from these two reactions, we prove in the main text that the proton implantation can induce oxidation, disproportionation, hydrogenation and dehydrogenation reactions. We therefore write the corresponding reactions for OT as:

Reaction 3: displacement damage

Reaction 4: oxidation

Reaction 5: disproportionation

$$2[\tfrac{1}{2}V_O]H:[\tfrac{1}{2}V_O] \rightarrow [\tfrac{1}{2}V_O]HO[\tfrac{1}{2}V_O] + [\square\tfrac{1}{2}V_O]H:[\tfrac{1}{2}V_O]$$

Reaction 6: hydrogenation

$$[\tfrac{1}{2}V_O]H:[\tfrac{1}{2}V_O] + H \rightarrow [\tfrac{1}{2}V_O]HH[\tfrac{1}{2}V_O]$$

Reaction 7: proton-assisted dehydrogenation

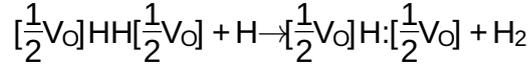

Based on the difference of formation energies, the reaction 5 is an energy increasing process. The reaction 3, 4, 5 and 7 all are energy decreasing processes, and we denote their sum as reaction 8:

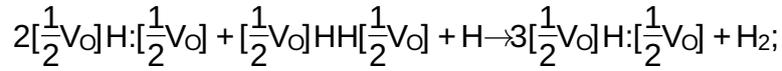

Reaction 8 is similar to R8 in the Brusellator-like model for $N_{it}$ in the main text. If we take $[\frac{1}{2}V_O]H:[\frac{1}{2}V_O]$, $[\frac{1}{2}V_O]HH[\frac{1}{2}V_O]$, $[\frac{1}{2}V_O][\frac{1}{2}V_O]$ as X, Y, C and D of the Brusellator model respectively, then the above reaction 1, 2, 6 and 8 corresponds to the reaction B1, B4, B3 and B2.